%% file: KASSGC_ARXIV.TEX
\documentclass[11pt,fleqn]{article}
\usepackage{amsmath,graphicx}

\input preamble.tex

\begin{document}
\twocolumn[

\Title{Relativistic algebra of space-time and algebrodynamics}

\Aunames{Vladimir V. Kassandrov\auth{a,1} and Joseph A. Rizcallah\auth{b,2}}

\Addresses{
\addr a {Institute of Gravitation and Cosmology, Peoples' Friendship
	University of Russia, Moscow, Russia}
\addr b {School of Education, Lebanese University, Beirut, Lebanon}
	}

\Abstract
  {We consider a manifestly Lorentz invariant form $\mathbb L$ of the biquaternion algebra and its generalization to the case of curved manifold. The conditions of $\mathbb L$-differentiability of $\mathbb L$-functions are formulated and considered as the primary equations for fundamental fields modeled with such functions. The exact form of the effective affine connection induced by $\mathbb L$-differentiability equations is obtained for the flat and curved cases. In the flat case, the integrability conditions of the latter leads to the self-duality of the corresponding curvature, thus ensuring that the source-free  Maxwell and $SL(2,\mathbb C)$ Yang-Mills equations hold on the solutions of the $\mathbb L$-differentiability equations.}

Keywords: {\em Quaternionic differentiability; covariantly constant field; self-duality}\\

\PACS{02.40.-k, 03.50.-z, 02.10.De} 

] 
\email 1 {vkassan@sci.pfu.edu.ru}
\email 2 {joeriz68@gmail.com}

\section{Lorentz invariant algebra of flat or curved space-time 
manifold}
\label{sec1}
  To construct a truly unified theory, one should rely on an exceptional geometrical structure. On the other hand, the diversity of geometries of different dimensions, topologies or differential structures does not allow for a trustful choice of the candidate for (an extended) space-time geometry. If, however, an {\it algebraic} structure is laid in the foundation of the theory, the situation becomes much better, for only a finite number of exceptional Lie groups or finite dimensional linear algebras exist, the latter being exhausted by complex numbers, quaternions and (non-associative) octonions.

Since the times of Hamilton, it is well known that the Euclidean structure of 3D physical space can be regarded as a direct consequence of the existence of the exceptional quaternion algebra with its group of automorphism $SO(3)$. A lot of effort has been made to relate the structure of Minkowski space-time $\bf M$ with the properties of complex quaternions ({\it biquaternions}) $\mathbb B$, whose symmetry group $SO(3,\mathbb C)$ is 2:1 isomorphic to the spinor Lorentz group $SL(2,\mathbb C)$, see, e.g., the review~\cite{Gsponer}. However, the $4\mathbb C$ dimension of this algebra corresponds to the structure of complexified space-time, in which $\bf M$ does not even constitute a subalgebra, but only a subspace. 

In~\cite{Grgin} an  interesting 4D algebra $\mathbb G$ has been proposed with a manifestly Lorentz covariant law of multiplication
\bearr
\label{Grgin}
(a\circ  b)^\mu = a^\mu (b^\rho e_\rho) + b^\mu (a^\rho e_\rho) - e^\mu (a^\rho b_\rho) 
\nnn \ \ \
\pm \imath \varepsilon^\mu_{. \nu\rho\lambda} a^\nu b^\rho e^\lambda,
\ear
where the ordinary notation using the Minkowski metric $\eta_{\mu\nu}$ is employed, and $\mu=0,1,2,3$. For the distinguished element $e=\{e^\mu\}$ of $\mathbb G$ the constraint $e^\mu e_\mu=1$ is imposed; then $e$ plays the role of the {\it unit} element: $a\circ e =e\circ a =a,~~ \forall a\in \mathbb G$. It is straightforward to verify that $\mathbb G$ is an {\it associative} algebra.

Surprisingly, the fact that $\mathbb G$ is isomorphic to the algebra of biquaternions $\mathbb B$ was overlooked in~\cite{Grgin}. Indeed, with the choice $e^\mu=\{1,0,0,0\}$ for the unit element, the law (\ref{Grgin}) reproduces the ordinary multiplication in $\mathbb B$. Precisely, for the basis vectors $\sigma_\mu$ equation (\ref{Grgin}) yields
\begin{equation}\label{base}
\sigma_\mu \circ \sigma_\nu = \sigma_\mu e_\nu + \sigma_\nu e_\mu - (e^\rho \sigma_\rho) \eta_{\mu\nu} \pm \imath \varepsilon^\rho_{. \mu\nu\lambda} \sigma_\rho e^\lambda, 
\end{equation}
with $\sigma_0=e$, and  
\begin{equation}\label{Pauli}
\sigma_a \circ \sigma_b = \delta_{ab} e \pm \imath \varepsilon_{abc} \sigma_c,
\end{equation}
for the three space-like basis vectors $\sigma_a,~a=1,2,3$, in full accord with the multiplication of the {\it Pauli matrices}. The different signs in (\ref{Grgin}) correspond to the left or right forms of the (bi)quaternion algebra. Under Lorentz boosts the unit element $e$ transforms as a 4-vector while the defining law (\ref{Grgin}) preserves its form. Moreover, 3-rotations are canonical automorphisms of $\mathbb G$.     
      
Representation (\ref{Grgin}) is very useful for generalizations to curved manifolds~\cite{Trish,Trish2}. To this end, we consider the {\it tetrad field}  $h_\mu^\alpha(x),~~\alpha,\beta, ...=0,1,2,3$ and the {\it local algebra} $\mathbb L$~\footnote{The concept of a local algebra has been introduced in~\cite[ch.2]{AD} and elaborated further in~\cite{Trish2}; the analogous concept of the so-called $\mathbb Q$-basis was considered in~\cite{Yefrem}.} defined by the basis vectors $\Sigma_\mu(x):=h_\mu^\alpha(x)\sigma_\alpha$ depending on the point of the manifold. The multiplication table (\ref{base}) for $\mathbb L$  takes then the obvious form 
\bearr
\label{Base}
\Sigma_\mu \circ \Sigma_\nu = \Sigma_\mu E_\nu + \Sigma_\nu E_\mu - (E^\rho \Sigma_\rho) g_{\mu\nu}  
\nnn \ \
\pm \imath \sqrt{-g}   \varepsilon^\rho_{. \mu\nu\lambda} \Sigma_\rho E^\lambda,  
\ear
in which the unit field $E_\mu(x):=h_\mu^\alpha e_\alpha$ satisfies $g^{\mu\nu}E_\mu E_\nu =1$, and a metric tensor of the manifold $g_{\mu\nu}:=h_\mu^\alpha h_\nu^\beta \eta_{\alpha\beta}$ ($g:=\vert g_{\mu\nu} \vert$) naturally arises.      

The existence of the local algebra (\ref{Base}) on a (complexified) 4D manifold 
requires, apart from the metric tensor, an algebro-geometrical structure -- the unit time-like 4-vector field ({\it U-field}) $E^\mu(x)$~\footnote{The situation resembles that in Weyl geometry which, together with the metric tensor, is defined by the nonmetricity 1-form field (identified by H. Weyl as the form of electromagnetic potentials).}. Its physical interpretation may be related to the flow of matter, etc. Below, we shall see that the properties of {$\mathbb L$-valued functions differentiable over $\mathbb L$} can impose strong restrictions on the metric and the U-field and, in a sense, determine the geometry of the manifold itself. First and foremost, however, they define a set of relativistic fields and guarantee the fulfillment of corresponding field equations.  

\section{Analysis and induced field dynamics on the relativistic algebra of space-time}

 In the algebrodynamical program (see, e.g.,~\cite{AD,GR95,YadPhys} and references therein) one considers some algebraic structure (``space-time algebra'') $\mathbb A$ which predetermines both the physical geometry and the equations of fundamental fields. Specifically, one should formulate the {\it differentiability conditions} for $\mathbb A$-valued functions, in close analogy to the Cauchy-Riemann conditions for holomorphic functions of complex variable. For quaternion-like, non-commutative yet associative, algebras appropriate conditions have been proposed in~\cite{AD,GR95} in the following Pfaffian form:
\begin{equation}\label{GCR}
dF = \Phi \circ dZ \circ \Psi, 
\end{equation}
in which  $F(Z), \Phi(Z), \Psi(Z)$ are the principal $\mathbb A$-function of $\mathbb A$-variable $Z$ and two auxiliary $\mathbb A$-functions (the so-called left and right {\it semi-derivatives}), respectively. $dF$ denotes the linear part of the increment (differential) of $F(Z)$, while $(\circ)$ -- the operation of multiplication in $\mathbb A$. 

Thus, an $\mathbb A$-function $F(Z)$ is called differentiable over $\mathbb A$ if its increment can be represented in the invariant form (\ref{GCR}), i.e. only through the operation of multiplication in $\mathbb A$. 

For commutative complex algebra the above condition reduces to $dF=(\Phi*\Psi)*dZ$ and, being written out in components, is equivalent to the Cauchy-Riemann equations. For real quaternions $\mathbb Q$ (\ref{GCR}) proves to be a condition of {\it conformity} of the corresponding mapping $Z\mapsto F(Z)$ in $\bf E^4$, in full analogy with the complex case. However, the class of such mappings is known to be very restricted, defined by 15 parameters only ({\it Liouville theorem}). Fortunately, upon complexification of $\mathbb Q$, that is, transition to the algebra of biquaternions $\mathbb B$, the class of solutions to (\ref{GCR}) substantially widens, on the account of elements $\Phi(Z)$ and/or $\Psi(Z)$ with {\it null} norm. 

Therefore, the $\mathbb B$-algebra (or, equivalently, the isomorphic $\mathbb G$-algebra) does constitute the basis of an algebrodynamical theory. $\mathbb B$-differentiable functions of the $\mathbb B$-variable should then be considered as the primary physical fields (together with the corresponding semi-derivatives) while the field equations are represented by the differentiability conditions (\ref{GCR}) or secondary constraints following from the latter (e.g., via successive differentiations, etc.). In this approach, {\it particles} are modeled as singularities of the $\mathbb B$-fields. 

In order to avoid problems with the complex extension of space-time (related to the complex $4\mathbb C$ structure of the vector space of $\mathbb B$-algebra), in our previous papers we restricted the coordinate space to the subspace with Minkowski metric. In the matrix representation of $\mathbb B$ this corresponds to Hermitian matrices,
\begin{equation}\label{hermit}
Z \mapsto X=X^+ = x^\mu \sigma_\mu, ~~\{x_\mu\} \in \mathbb R
\end{equation}
while the components of the fundamental fields $F(X),\Phi(X)$ and $\Psi(X)$ are generally assumed to be complex-valued. Finally, we regard the $\mathbb B$-differentiability conditions on the Minkowski coordinate subspace 
\begin{equation}\label{GCRM}
dF = \Phi(X)\circ dX  \circ \Psi(X)
\end{equation}
as the only constraints to determine the corresponding functions-fields and their singular locus, which is identified with particle-like formations.   

Using the $SL(2,\mathbb C)$ matrix representation of the $\mathbb B$-algebra, it was 
proved~\cite{AD,GR95} that any matrix component $\varphi=F_A^B(x)$ of the $\mathbb B$-field satisfies, in view of (\ref{GCRM}), the complex eikonal equation
\begin{equation}\label{eik}
\eta^{\mu\nu}\partial_\mu \varphi \partial_\nu \varphi = 0, 
\end{equation}
instead of the linear Laplace equation for complex functions. In general, the system of PDE corresponding to (\ref{GCRM}) is Lorentz invariant and {\it nonlinear} 
(the latter is a direct consequence of the non-commutativity of $\mathbb B$-algebra). 

For the most important~\cite{PhDRiz} case when $\Psi(x)\equiv F(X)$ (equivalently, one can take $\Phi(X)\equiv F(X)$), after spinor splitting of (\ref{GCRM}), we obtain
\begin{equation}\label{GSE}
d\xi = \Phi dX \xi,
\end{equation} 
for the 2-spinor $\xi=\{\xi_A(x)\}$ and complex 4-vector $\Phi=\{\Phi_{AB^\prime}(x)\}$ fields ($A,A^\prime,...=0,1$). Both $\xi_A(x)$ and $\Phi_{AB^\prime}(x)$ can be found from the overdetermined structure of the system of equations (\ref{GSE}). 

In particular, the integrability conditions of (\ref{GSE}) read
\begin{equation}\label{integrab}
dd\xi = R\xi = 0, ~~R:=(d\Phi - \Phi dX \Phi)\wedge dX, 
\end{equation}
where the matrix-valued 2-form $R$ can be regarded as the {\it curvature 2-form} of the matrix-valued {\it connection 1-form} $\Omega:= \Phi dX$ entering the initial equations (\ref{GSE}). The latter can thus be interpreted as the conditions for 2-spinor field $\xi(X)$ be {\it covariantly constant} w.r.t. the complex affine connection $\Omega$, 
\begin{equation}\label{CCVF}
d\xi - \Omega \xi = 0.
\end{equation}     

Let us now return to the integrability conditions (\ref{integrab}). Since the spinor $\xi(x)$ is not arbitrary, the curvature $R$ is not null. Thus, the $\mathbb B$-differentiability conditions {\it dynamically}, on the flat Minkowski background, define a non-trivial geometric structure -- the {\it complex curved 4D space} with affine connection $\Omega$. Moreover, it has been obtained in~\cite{GR95,KasRiz} that the spinor components can be eliminated from (\ref{integrab}), and the curvature $R$ turns out to be {\it self-dual} on the solutions of (\ref{GSE}). Specifically, one obtains from (\ref{integrab}):
\begin{equation}\label{constraint}
(\vec {\bf R})_a:=R_{oa}+\frac{\imath}{2} \varepsilon_{abc}R_{bc} = 0,
\end{equation}
 with the following structure of the self-dual part of curvature:
\begin{equation}\label{selfdual}
\vec {\bf R} = \vec P +D \vec \sigma -\imath \vec P \times \vec \sigma, 
\end{equation}
 where the quantities $\vec P:=\vec E + \imath \vec H$ and $D$ are defined through the components of the 4-vector field $\Phi=A^\mu(x) \sigma_\mu$ as follows:
\bearr
\label{pot}
\vec E:=-\partial_o \vec A -\nabla A_o,~~\vec H:=\nabla \times \vec A,
\nnn \ \ 
D:=\partial_\mu A^\mu +2 A_\mu A^\mu,
\ear
and represent, consequently, the components of 4-potentials and field strengths of an effective complex electromagnetic field. Now, from the full self-duality condition (\ref{constraint}), the self-duality of electromagnetic field follows immediately,
\begin{equation}\label{EMselfdual}
\vec E + \imath \vec H =0,
\end{equation}
together with the ``inhomogeneous Lorentz condition'' $D=0$. 

In turn, the complex self-duality condition (\ref{EMselfdual}) guarantees the source-free Maxwell equations, separately for the (mutually dual) real and imaginary parts of the electromagnetic fields (\ref{pot}). Moreover, for two independent components $\psi(x)$ of the potential matrix $\Phi(X)$ (the other two can always be nullified by a gauge transformation) {\it the 2-spinor Weyl equation holds for any solution of (\ref{GSE})}~(for details, see~\cite{Sing}). 
Remarkably, the $SL(2,\mathbb C)$ Yang-Mills fields can be also defined through the same matrix field $\Phi(X)$ and, on the solutions of (\ref{GSE}),  satisfy the corresponding source-free equations; for proofs and details we refer the reader to~\cite{GR95,KasRiz}).   

From a 4-vector perspective, the principal equation (\ref{GSE}) reads 
\begin{equation}\label{GSEV}\
\partial_\nu F = \Phi \circ \sigma_\nu \circ F, 
\end{equation}
or, in components,
\begin{equation}\label{4GSE}
\partial_\nu F^\rho  \sigma_\rho = A^\mu F^\rho \sigma_\mu \circ \sigma_\nu \circ  \sigma_\rho. 
\end{equation}  
Using (\ref{base}) to evaluate the product in the r.h.s., we find 
\begin{equation}\label{4CCF}
\partial_\nu F^\rho =\Gamma_{\nu\mu}^\rho F^\mu,  
\end{equation}
with the connection $\Gamma_{\nu\mu}^\rho $ being
\bearr
\label{4conn}
\Gamma_{\nu\mu}^\rho = \delta^{\rho}_{\nu}A^\alpha(2e_\mu e_\alpha - \eta_{\mu\alpha})- A_{\nu}\eta^{\beta \rho}(2e_\mu e_\beta - \eta_{\mu\beta}) 
\nnn \ \
\pm \imath \{\epsilon^{\rho}_{. \alpha \nu \gamma}e_{\mu} +\epsilon^{\rho}_{. \alpha \mu \gamma}e_{\nu} -\epsilon^{\rho}_{. \nu \mu \gamma}e_{\alpha} + \epsilon_{\alpha \nu \mu \gamma}e^{\rho}\}e^\gamma A^\alpha
\nnn \ \
+ A^\rho(2e_\nu e_\mu - \eta_{\nu\mu}).
\ear
In this form, equations (\ref{4CCF}) and (\ref{4conn}) define a covariantly constant 4-vector field and can be readily generalized to curved metric-affine  spaces. Specifically, instead of (\ref{GCR}) we now can deal with the $\mathbb L$-differentiability conditions of the form (in the principal case of $\Psi(Z)\equiv F(Z)$): 
\begin{equation}\label{RiemannGCR}
DF =\Phi(Z)\circ dX \circ F(Z), 
\end{equation}
with $DF$ being the covariant differential w.r.t. the metric $g_{\mu\nu}$ (that is, w.r.t. the Levi-Civita connection $\gamma$)~\footnote{A more sophisticated and mathematically substantiated approach to the generalization of differentiability conditions (\ref{GCR}) has been elaborated in~\cite{Trish2}.}. Again, (\ref{RiemannGCR}) can be interpreted as the conditions for 4-vector $F(Z)$ be covariantly constant w.r.t the connection
\begin{equation}\label{fullconn}
\Gamma=\gamma + G, 
\end{equation} 
where $G:=\{G_{\nu\mu}^\rho\}$ is the connection of the form (\ref{4conn}) generalized, in a natural way, to the case of local $\mathbb L$-algebra (\ref{Base}):
\bearr
\label{Wconn}
G_{\nu\mu}^\rho = \delta^{\rho}_{\nu}A^\alpha(2E_\mu E_\alpha - g_{\mu\alpha}) - A_{\nu}g^{\beta \rho}(2E_\mu E_\beta - g_{\mu\beta})   
\nnn \ \
\pm \imath\sqrt{-g}\{\epsilon^{\rho}_{. \alpha \nu \gamma}E_{\mu} +\epsilon^{\rho}_{. \alpha \mu \gamma}E_{\nu} -\epsilon^{\rho}_{. \nu \mu \gamma}E_{\alpha}  
\nnn \ \
+\epsilon_{\alpha \nu \mu \gamma}E^{\rho}\}E^\gamma A^\alpha + A^\rho(2E_\nu E_\mu - g_{\nu\mu}).
\ear

It follows from (\ref{Wconn}) that the generalized $\mathbb L$-connection is not symmetric in the lower indexes and thus possesses {\it torsion} of a rather specific form. On the other hand, for the covariant derivative of metric tensor w.r.t. the connection (\ref{Wconn}), one has:
\begin{equation}\label{Weyl}
\nabla_\rho g_{\mu\nu} = -2 g_{\mu\nu} \tilde A_\rho,
\end{equation}
with $\tilde A_\rho:= (2E_\rho E_\lambda - g_{\rho\lambda})A^\lambda$. Thus, the considered connection (\ref{Wconn}) possesses {\it nonmetricity} of the Weyl type, and we are effectively dealing with a dynamically induced, by the primary algebraic structure, {\it Weyl-Cartan manifold}~\cite{GR95}. 

It is worth noting that the structure of $\mathbb L$-connection (\ref{Wconn}) and the covariant derivative of the metric (\ref{Weyl}) essentially contain the effective metric tensor 
\begin{equation}\label{newmetric}
\tilde g_{\rho\lambda}:= 2E_\rho E_\lambda - g_{\rho\lambda}
\end{equation}
which, rather surprisingly, in the ``flat'' limit of $\mathbb B$-algebra and connection (\ref{4conn}) reduces to the metric of 4D {\it Euclidean} (!) space 
and preserves its Euclidean structure upon generalization to (\ref{Wconn}). 

Moreover, a third metric $g^*_{\mu\nu}(X)$ defined algebraically via the {\it structure functions} $C^\rho_{\mu\nu}(X)$ of the local $\mathbb L$-algebra and so invariant under the automorphisms of $\mathbb L$ {\it turns out to coincide with (\ref{newmetric}}) (at least in the case $g=\vert g_{\mu\nu}\vert=\vert \eta_{\mu\nu}\vert =-1$):
\begin{equation}\label{autometric}
g^*_{\mu\nu}:= \frac{1}{4}C^\beta_{\mu\alpha} C^\alpha_{\nu\beta}\equiv \tilde g_{\mu\nu}.
\end{equation}
The possible meaning of this effective metric and the consequences of the obtained remarkable coincidence deserve further investigation!

In light of all the above, it is not unreasonable to assume that the integrability conditions of the $\mathbb L$-differentiability equations (\ref{RiemannGCR}) might impose restrictions not only on the vector field $A_\mu$, but also on the metric $g_{\mu \nu}$ as well as the unit vector field $E_\mu$. However, the task of analyzing these conditions and the equations they yield for the metric and fields is left for future work.    

\section{Conclusion}
We study a manifestly covariant form of the biquaternion algebra $\mathbb B$, which coincides with the relativistic ring extension proposed in~\cite{Grgin}. We consider the $\mathbb B$-differentiability conditions of $\mathbb B$-valued functions and suggest to regard them as the fundamental generating system for the set of relativistic physical fields. In the most important case, these conditions admit a geometrical interpretation as those defining a covariantly constant vector field, w.r.t. an affine connection of a very specific form. Their integrability conditions lead then to the self-duality of the corresponding curvature, which in turn yields Maxwell, Yang-Mills and Weyl equations for the associated fields. It is hoped that $\mathbb L$-generalization of the $\mathbb B$-algebra and $\mathbb B$-differentiability equations to Riemannian or general metric-affine space-times will enable us, in addition, to obtain sufficient constraints on the connection, metric and unit $U$-field and thus, in a purely algebraic way, determine the physical geometry.

\end{document}

%% file: preamble.tex
\usepackage{amsfonts,amssymb,cite}
\usepackage{epsf,graphicx}

\def\noi{\noindent}

\newcommand{\Title}[1]{\noi {{\Large\bf #1}}\\[1ex]}

\def\Aunames#1{\noi{\bf #1}}
\def\auth#1{${}^{#1}$}
\def\Addresses#1{\medskip\noi \protect
	\begin{description}\itemsep -3pt {\it #1} \end{description}}
\def\addr#1#2{\item[${}^{#1}$]{\it #2}}

\newcommand{\Abstract}[1]{\vskip 2mm \begin{center}
        \parbox{16.4cm}{\small\noi #1} \end{center}\medskip}
\newcommand{\PACS}[1]{\begin{center}{\small PACS: #1}\end{center}}

\def\email#1#2{\footnotetext[#1]{e-mail: #2}\addtocounter{footnote}{1}}

\def\nqq{\hspace*{-2em}}

\def\Jl#1#2{#1 {\bf #2},\ }

\def\ApJ#1 {\Jl{Astroph. J.}{#1}}
\def\CQG#1 {\Jl{Class. Quantum Grav.}{#1}}
\def\DAN#1 {\Jl{Dokl. AN SSSR}{#1}}
\def\GC#1 {\Jl{Grav. Cosmol.}{#1}}
\def\GRG#1 {\Jl{Gen. Rel. Grav.}{#1}}
\def\JETF#1 {\Jl{Zh. Eksp. Teor. Fiz.}{#1}}
\def\JETP#1 {\Jl{Sov. Phys. JETP}{#1}}
\def\JHEP#1 {\Jl{JHEP}{#1}}
\def\JMP#1 {\Jl{J. Math. Phys.}{#1}}
\def\NPB#1 {\Jl{Nucl. Phys. B}{#1}}
\def\NP#1 {\Jl{Nucl. Phys.}{#1}}
\def\PLA#1 {\Jl{Phys. Lett. A}{#1}}
\def\PLB#1 {\Jl{Phys. Lett. B}{#1}}
\def\PRD#1 {\Jl{Phys. Rev. D}{#1}}
\def\PRL#1 {\Jl{Phys. Rev. Lett.}{#1}}

\def\lal{&&\nqq {}}

\def\beq{\begin{equation}}
\def\eeq{\end{equation}}
\def\bear{\begin{eqnarray}}
\def\bearr{\begin{eqnarray} \lal}
\def\ear{\end{eqnarray}}
\def\earn{\nonumber \end{eqnarray}}

\def\nnn{\nonumber\\ \lal }

